\documentclass[preprint,1p,sort&compress,times]{elsarticle}

\journal{Wave Motion}
\usepackage[fleqn]{amsmath}
\allowdisplaybreaks
\usepackage{amsfonts}
\usepackage{amssymb}
\usepackage{mathrsfs}
\usepackage{amsthm}
\usepackage{graphicx}
\usepackage{bm}
\usepackage{booktabs}

\usepackage{xcolor}

\newproof{pf}{Proof}

\begin{document}
	
	\begin{frontmatter}

		\title{
			Plane-strain waves in nonlinear elastic solids with softening
		}
		
		\author[lma]{Harold~Berjamin}
		\author[lma]{Bruno~Lombard}
		\ead{lombard@lma.cnrs-mrs.fr}
		\address[lma]{Aix-Marseille Univ, CNRS, Centrale Marseille, LMA, Marseille, France}
		
		\author[m2p2]{Guillaume~Chiavassa}
		\address[m2p2]{Centrale~Marseille, CNRS, Aix-Marseille Univ, M2P2~UMR~7340, 13451~Marseille~Cedex~20, France}
		
		\author[iusti]{Nicolas~Favrie}
		\address[iusti]{Aix-Marseille Univ, UMR~CNRS~7343, IUSTI, Polytech~Marseille, 13453~Marseille~Cedex~13, France}
		
		\cortext[cor]{Corresponding author}
		
		\begin{abstract}
			
			Propagation of elastic waves in damaged media (concrete, rocks) is studied theoretically and numerically. Such materials exhibit a nonlinear behavior, with long-time softening and recovery processes (slow dynamics). A constitutive model combining Murnaghan hyperelasticity with the slow dynamics is considered, where the softening is represented by the evolution of a scalar variable. The equations of motion in the Lagrangian framework are detailed. These equations are rewritten as a nonlinear hyperbolic system of balance laws, which is solved numerically using a finite-volume method with flux limiters. Numerical examples illustrate specific features of nonlinear elastic waves, as well as the effect of the material's softening. In particular, the generation of solitary waves in a periodic layered medium is illustrated numerically.
			
		\end{abstract}
		

	\end{frontmatter}

	\section{Introduction}\label{sec:Intro}
	
	Geomaterials such as rocks and concrete exhibit nonlinear features at small strains $\varepsilon \sim 10^{-6}$. In particular, longitudinal resonance experiments reveal the generation of higher-order harmonics. Besides this phenomenon, the material exhibits also a softening with increasing strain amplitudes \cite{johnson96b,remillieux16}. After the forcing is stopped, the material recovers gradually its initial stiffness. This softening/recovery process is not instantaneous. The transient regime, named ``slow dynamics'' in the corresponding literature \cite{tencate96,tencate11,riviere13,haupert19}, has a characteristic time much larger than the period of the forcing.
	
	To describe nonlinear elastic behavior, the finite-strain theory is a self-consistent framework. Various constitutive laws of isotropic hyperelastic material express the stress as a function of a strain tensor, and are compatible with Hooke's law in the infinitesimal strain limit \cite{ogdenElast84,holzapfel00,norris98}. Among them, the Murnaghan law \cite{murnaghan37} is frequently used to describe acoustic nonlinearity in rocks and concrete, and values of the parameters can be found in the literature \cite{johnson96a,winkler96,payan09}. However, hyperelasticity does not account for the long-time relaxation of elastic constants.
	
	Several models from the literature describe the softening of the material by a dependence of the elastic constants on a scalar variable $g$ \cite{lyakhovsky97a,vakhnenko04,berjamin17a}. In the present study, the internal-variable model \cite{berjamin17a} is used, where the stress issued from Murnaghan's law is multiplied by $1-g$. Hence, $g \equiv 0$ corresponds to the hyperelastic case, and the material softens as $g$ increases. The evolution of $g$ is governed by a first-order differential equation compatible with the principles of thermodynamics. Both softening processes $\dot g\geqslant 0$ and recovery processes $\dot g\leqslant 0$ are possible \cite{berjamin17a}, accordingly to the experimental observations.
	
	In recent works \cite{berjamin18b}, the corresponding one-dimensional equations of motion have been solved numerically using a finite-volume method with flux limiters \cite{leveque02}. The numerical method has been validated with respect to reference solutions, and qualitative agreement with experimental observations has been obtained. The present article describes a similar method in the two-dimensional plane-strain case. The numerical method is well-suited to the computation of nonlinear waves in the Lagrangian framework, and it can be used for various hyperelastic material models (cf. the related study \cite{boumatar12} and references therein).
	
	The article is organized as follows. In Section~\ref{sec:Eq}, the equations of motion are detailed, which are rewritten as a nonlinear hyperbolic system of balance laws in two space variables. The numerical method is presented in Section~\ref{sec:Num}. Section~\ref{sec:NumExp} shows 2D numerical results illustrating the nonlinear wave propagation in elastic solids with softening. The case of a homogeneous medium is considered, as well as the case of a periodic layered medium with solitary wave solutions.
	
	
	\section{Governing equations}\label{sec:Eq}
	
	\subsection{Lagrangian hyperelasticity with softening}
	
	We consider an homogeneous continuum. A particle initially located at some position $\bm{x}_0$ of the reference configuration moves to a position $\bm{x}_t$ of the deformed configuration. The deformation gradient is a second-order tensor defined by (see e.g. \cite{ogdenElast84,holzapfel00,drumheller98,norris98})
	\begin{equation}
		\bm{F} = \mathbf{grad}\, \bm{x}_t = \bm{I} + \mathbf{grad}\, \bm{u}\, ,
		\label{F}
	\end{equation}
	where $\bm{u} = \bm{x}_t - \bm{x}_0$ denotes the displacement field and $\mathbf{grad}$ is the gradient with respect to the material coordinates $\bm{x}_0$ (Lagrangian gradient). In the reference configuration, the deformation gradient \eqref{F} is equal to the metric tensor $\bm{I}$. Here, the Euclidean space is described by an orthonormal basis $(\bm{e}_1,\bm{e}_2,\bm{e}_3)$ and a Cartesian coordinate system $(O,x,y,z)$, so that the matrix of the coordinates of $\bm I$ is the identity matrix.
	
	The Lagrangian representation of motion is used. Hence, the material time derivative $\dot{\bm{F}} = \partial_t\bm{F}$ of the deformation gradient satisfies
	\begin{equation}
		\dot{\bm{F}} = \mathbf{grad}\,\bm{v} \, ,
		\label{FDotGen}
	\end{equation}
	where $\bm{v} = \dot{\bm{u}}$ is the velocity field. The conservation of mass implies ${\rho_0}/{\rho} = \det \bm{F}$,
	where $\rho$ denotes the mass density in the deformed configuration, and $\rho_0$ denotes the mass density in the reference configuration. Self-gravitation and heat conduction are neglected, so that the motion is driven by the conservation of momentum
	\begin{equation}
		\rho_0 \dot{\bm{v}} = \mathbf{div}\,\bm{P} + \bm{f}^v\, ,
		\qquad\text{where}\qquad
		\bm{P} = \det(\bm{F})\,\bm{\sigma}\cdot\bm{F}^{-\top}
		\label{ConsMom}
	\end{equation}
	is the first Piola--Kirchhoff stress tensor, and $\mathbf{div}$ denotes the divergence with respect to the material coordinates $\bm{x}_0$.
	The Cauchy stress tensor $\bm{\sigma} = (\det \bm{F})^{-1} \bm{P}\cdot\bm{F}^\top\! = \bm{\sigma}^\top\!$ is detailed later on through a specification of $\bm{P}$. The term $\bm{f}^v$ is an external volume force applied to the material.
	
	In hyperelasticity, the only variables of state are the specific entropy and a strain tensor. Here, the Green--Lagrange strain tensor $\bm{E} = \frac{1}{2} (\bm{F}^\top\!\cdot\bm{F} - \bm{I})$ is used, i.e.
	\begin{equation}
		\bm{E} = \frac{1}{2} \left( \mathbf{grad}\, \bm{u} + \mathbf{grad}^\top \bm{u} + \mathbf{grad}^\top \bm{u}\cdot \mathbf{grad}\, \bm{u} \right) .
		\label{E}
	\end{equation}
	An internal variable $g \in [0,1[$ accounting for the softening of the material is added to the previous list of variables of state.
	We define the internal energy by unit of reference volume as \cite{berjamin17a}
	\begin{equation}
		\rho_0 e = \left(1-g\right) W(\bm{E}) + \Phi(g) \, ,
		\label{InternalEnergy}
	\end{equation}
	where $e$ is the specific internal energy. In \eqref{InternalEnergy}, the potential $W(\bm{E})$ is the strain energy function of Murnaghan's law \cite{murnaghan37}
	\begin{equation}
		W(\bm{E}) = \frac{\lambda + 2 \mu}{2} {E_\text{I}}^2 - 2 \mu {E_\text{II}} + \frac{\mathfrak{l} + 2 \mathfrak{m}}{3} {E_\text{I}}^3 - 2\mathfrak{m}{E_\text{I}}{E_\text{II}} + \mathfrak{n}{E_\text{III}} \, ,
		\label{EnergyStrain}
	\end{equation}
	which is expressed as a function of the strain invariants ${E_\text{I}} = \text{tr}\,\bm{E}$, ${E_\text{II}} = \frac{1}{2}\big( (\text{tr}\,\bm{E})^2 - \text{tr}(\bm{E}^2) \big)$, and ${E_\text{III}} = \det \bm{E}$.
	The constants $\lambda$, $\mu$ are the Lam{\'e} parameters and the constants $\mathfrak{l}$, $\mathfrak{m}$, $\mathfrak{n}$ are the Murnaghan coefficients (third-order elastic constants). In the case where the Murnaghan coefficients in \eqref{EnergyStrain} equal zero, the strain energy of the Saint~Venant--Kirchhoff model is recovered.
	The potential $\Phi(g)$ represents a storage energy. Basic requirements are the convexity of $\Phi$ and $\Phi'(0)=0$, where $\Phi'$ is the derivative of $\Phi$ (see \cite{berjamin17a}). Moreover, an asymptote at $g=1$ is introduced to avoid the destruction of the material. A suitable expression is
	\begin{equation}
		\Phi(g) = -\frac{1}{2}\gamma \ln (1-g^2)\, ,
		\label{EnergyStorage}
	\end{equation}
	where $\gamma > 0$ is an energy per unit volume. The energy \eqref{EnergyStorage} is quadratic $\Phi(g) \simeq \frac{1}{2}\gamma g^2$ in the limit $g\to 0$.
	
	Under the previous assumptions, the simplest set of constitutive equations with softening which is thermodynamically admissible reads \cite{berjamin17a}
	\begin{equation}
		\bm{P} = \left(1-g\right) \bm{F}\cdot\frac{\partial W}{\partial\bm{E}}\,
		,
		\qquad\text{and}\qquad
		\tau_1\dot{g}  = W(\bm{E}) - \Phi'(g)\, ,
		\label{Constitutive}
	\end{equation}
	where $\tau_1>0$ in J\,m\textsuperscript{$-3$}\,s is a material parameter. The first equation in \eqref{Constitutive} is the mechanical constitutive law, which reduces to the case of hyperelasticity if $g \equiv 0$. According to the second equation in \eqref{Constitutive} which governs the evolution of the variable $g$, the classical theory of nonlinear elastodynamics is recovered if $\tau_1 \to {+\infty}$.
	Similarly to \cite{berjamin17a,berjamin18b}, we rewrite the tensor derivative $\partial W/\partial\bm E$ of the strain energy using the invariants' tensor derivatives:
	\begin{equation}
		\frac{\partial W}{\partial\bm{E}} = \alpha_0 \bm{I} + \alpha_1 \bm{E} + \alpha_2 \bm{E}^2 ,
		\label{PK2}
	\end{equation}
	where
	\begin{equation}
		{\addtolength{\jot}{0.2em}
		\begin{aligned}
			\alpha_0 & = \frac{\partial W}{\partial{E_\mathrm{I}}} + E_\mathrm{I}\frac{\partial W}{\partial{E_\mathrm{II}}} + E_\mathrm{II}\frac{\partial W}{\partial{E_\mathrm{III}}}\! && = \lambda E_\mathrm{I} + \mathfrak{l} {E_\mathrm{I}}^2 - (2\mathfrak{m}-\mathfrak{n}) E_\mathrm{II} \, , \\
			\alpha_1 & = -\frac{\partial W}{\partial{E_\mathrm{II}}} - E_\mathrm{I}\frac{\partial W}{\partial{E_\mathrm{III}}}\! && = 2\mu + (2\mathfrak{m}-\mathfrak{n}) E_\mathrm{I} \, , \\
			\alpha_2 & = \frac{\partial W}{\partial{E_\mathrm{III}}}\! && = \mathfrak{n} \, .
		\end{aligned}}
		\label{PK2coeffs}
	\end{equation}
	In the next subsection, we detail the case of plane strain. The case of uniaxial strain is addressed in \cite{berjamin18b}.
	
	\subsection{The plane-strain assumption}
	
	The displacement field $\bm{u}$ is independent of $z$, and its component $u_3$ along $\bm{e}_3$ is zero. In the basis of unit tensors $(\bm{e}_i \otimes \bm{e}_j)_{1\leqslant i,j\leqslant 3}$, the matrix of coordinates of the displacement gradient $\mathbf{grad}\,\bm{u}$ is therefore
	\begin{equation}
		\begin{pmatrix}
			u_{i,j}
		\end{pmatrix} = {
		\renewcommand{\arraystretch}{1.2}
		\begin{pmatrix}
			u_{1,1} & u_{1,2} & 0 \\
			u_{2,1} & u_{2,2} & 0 \\
			0 & 0 & 0 
		\end{pmatrix}}\, .
		\label{GradUcomp2D}
	\end{equation}
	Using the Einstein notation with indices in $\lbrace 1 ,2\rbrace$, the coordinates of the Green--Lagrange tensor \eqref{E} write $E_{ij} = \frac{1}{2} \big( u_{i,j} + u_{j,i} + u_{p,i}u_{p,j}\big)$. Its invariants in \eqref{EnergyStrain} are $E_\text{I} = E_{nn}$, $E_\text{II} = \frac{1}{2} \big( {E_\text{I}}^2 - E_{ij}E_{ij}\big) = \epsilon_{ij} E_{1i}E_{2j}$ and $E_\text{III} = 0$, where $\epsilon_{ij}$ is the Levi-Civita symbol of $\mathbb{R}^2$.
	The Cayley--Hamilton theorem applied to the restriction of $\bm E$ to $\mathbb{R}^2\times \mathbb{R}^2$ reads $E_{im}E_{mj} - E_\text{I} E_{ij} + E_\text{II}\delta_{ij} = 0$, where $\delta_{ij}$ is the Kronecker delta. Hence, the expression \eqref{PK2} of $\partial W/\partial \bm{E}$ becomes
	\begin{equation}
		\frac{\partial W}{\partial E_{ij}} = \tilde{\alpha}_0 \delta_{ij} + \tilde{\alpha}_1 E_{ij}
		\label{PK22D}
	\end{equation}
	in the basis of unit tensors $(\bm{e}_i \otimes \bm{e}_j)_{1\leqslant i,j\leqslant 2}$, where
	\begin{equation}
		{\addtolength{\jot}{0.2em}
		\begin{aligned}
			\tilde{\alpha}_0 &= \alpha_0 - \alpha_2 E_\mathrm{II}\!  &&= \lambda E_\text{I} + \mathfrak{l} {E_\text{I}}^2 - 2\mathfrak{m} E_\text{II} \, , \\
			\tilde{\alpha}_1 &= \alpha_1 + \alpha_2 E_\mathrm{I}\! &&= 2\left(\mu + \mathfrak{m} E_\text{I} \right) .
		\end{aligned}}
		\label{PK22Dcoeffs}
	\end{equation}
	The components $P_{ij}$ of the first Piola--Kirchhoff stress tensor $\bm P$ in \eqref{Constitutive} are therefore
	\begin{equation}
		P_{ij} = (1-g) \left(\delta_{im} + u_{i,m}\right) \left(\tilde{\alpha}_0 \delta_{mj} + \tilde{\alpha}_1 E_{mj} \right)
		\label{PK12D}
	\end{equation}
	under the plane strain assumption, which does not depend upon the third Murnaghan coefficient $\mathfrak{n}$.
	
	When the geometric nonlinearities are negligible, the Green--Lagrange strain tensor \eqref{E} is linearized with respect to $\mathbf{grad}\,\bm{u}$, i.e. $\bm{E} \simeq \frac{1}{2} \big(\mathbf{grad}\,\bm{u} + \mathbf{grad}^\top\bm{u}\big) = \bm{\varepsilon}$ reduces to the infinitesimal strain tensor. The coordinates $E_{ij}$ of $\bm E$ are replaced by the coordinates $\varepsilon_{ij} = \frac{1}{2} \big( u_{i,j} + u_{j,i} \big)$ of $\bm{\varepsilon}$.
	Moreover, the first Piola--Kirchhoff stress tensor $\bm{P}$ is linearized with respect to $\mathbf{grad}\,\bm{u}$ too, i.e. $\bm{F}\cdot\partial W/\partial \bm{E} \simeq \partial W/\partial \bm{\varepsilon}$. Hence, the equation $P_{ij} = (1-g)\, \big( \tilde{\alpha}_0 \delta_{ij} + \tilde{\alpha}_1 \varepsilon_{ij} \big)$ replaces \eqref{PK12D}. Under this assumption, linear elastodynamics is recovered if $g \equiv 0$ (i.e., $\tau_1\to{+\infty}$ in \eqref{Constitutive}), and if the Murnaghan coefficients $\mathfrak{l}$, $\mathfrak{m}$ in \eqref{PK22Dcoeffs} are zero (see \ref{app:Eigenvects} for details).
	
	\subsection{System of balance laws}\label{subsec:Cons}
	
	Under the plane-strain assumption, the equations of motion \eqref{FDotGen}-\eqref{ConsMom}-\eqref{Constitutive} are rewritten as a two-dimensional system of balance laws
	\begin{equation}
		\partial_t \mathbf{q} + \partial_x \mathbf{f}(\mathbf{q}) + \partial_y \mathbf{g}(\mathbf{q}) = \mathbf{r}(\mathbf{q}) + \mathbf{s} \, ,
		\label{SystCons}
	\end{equation}
	where $\mathbf{q} = (u_{1,1},u_{1,2},u_{2,1},u_{2,2},v_1,v_2,g)^\top\!$ is the vector of unknowns. The expressions of the flux functions, the relaxation function and the source term are
	\begin{equation}
		\mathbf{f}(\mathbf{q}) = -
		{\renewcommand{\arraystretch}{1.2}
		\begin{pmatrix}
			v_1\\
			0\\
			v_2\\
			0\\
			P_{11}/\rho_0\\
			P_{21}/\rho_0\\
			0
		\end{pmatrix}}
		\, ,\quad
		\mathbf{g}(\mathbf{q}) = -
		{\renewcommand{\arraystretch}{1.2}
			\begin{pmatrix}
			0\\
			v_1\\
			0\\
			v_2\\
			P_{12}/\rho_0\\
			P_{22}/\rho_0\\
			0
			\end{pmatrix}}
		\, ,\quad
		\mathbf{r}(\mathbf{q}) = \frac{1}{\tau_1}
		{\renewcommand{\arraystretch}{1.2}
		\begin{pmatrix}
			0\\
			0\\
			0\\
			0\\
			0\\
			0\\
			W-\Phi'(g)
		\end{pmatrix}}
		\, ,\quad
		\mathbf{s} = \frac{1}{\rho_0}
		{\renewcommand{\arraystretch}{1.2}
		\begin{pmatrix}
			0\\
			0\\
			0\\
			0\\
			\bm{f}^v\!\cdot\bm{e}_1\\
			\bm{f}^v\!\cdot\bm{e}_2\\
			0
		\end{pmatrix}}
		\, .
	\label{SystConsFlux}
	\end{equation}
	In \eqref{SystConsFlux}, the Piola--Kirchhoff stress components $(P_{ij})_{1\leqslant i,j\leqslant 2}$ depend on $(u_{i,j})_{1\leqslant i,j\leqslant 2}$ and $g$ according to \eqref{PK12D}. The strain energy $W$ depends on $(u_{i,j})_{1\leqslant i,j\leqslant 2}$ according to \eqref{EnergyStrain}. 
	
	The Jacobian matrix of the flux component $\mathbf{f}$ along the $x$-axis is
	\begin{equation}
		\mathbf{f}'(\mathbf{q}) = -
			{\renewcommand{\arraystretch}{1.2}
			\begin{pmatrix}
				& & & & 1 & 0 & 0\\
				& & & & 0 & 0 & 0\\
				& & & & 0 & 1 & 0\\
				& & & & 0 & 0 & 0\\
				Q_{1111} & Q_{1112} & Q_{1121} & Q_{1122} & 0 & 0 & G_{11}\\
				Q_{2111} & Q_{2112} & Q_{2121} & Q_{2122} & 0 & 0 & G_{21}\\
				0 & 0 & 0 & 0 & 0 & 0 & 0
			\end{pmatrix}}\, ,
	\label{SystConsJacobiX}
	\end{equation}
	where only three strips are displayed (everywhere else, the coefficients in the matrix are zero). The expression of the coefficients $Q_{ijk\ell}$ in \eqref{SystConsJacobiX} defined by $\rho_0 Q_{ijk\ell} = \partial P_{ij}/\partial u_{k,\ell}$ is detailed in the \ref{app:Eigenvects}, as well as the expression of the coefficients $G_{ij}$ defined by $\rho_0 G_{ij} = \partial P_{ij}/\partial g$. A similar Jacobian matrix $\mathbf{g}'(\mathbf{q})$ is obtained for the flux component $\mathbf{g}$ along the $y$-axis. These matrices are diagonalized in the \ref{app:Eigenvects}. The spectrum of both matrices has the form $\lbrace -c_P, c_P, -c_S, c_S,0,0,0\rbrace$. In the case of Murnaghan hyperelasticity, the eigenvalues $c_P$, $c_S$ can be complex \cite{berjamin18b}, so that the system \eqref{SystCons}-\eqref{SystConsFlux} is not unconditionally hyperbolic (see e.g. \cite{favrie14b} for discussions on hyperbolicity in hyperelasticity). Here, we restrict ourselves to configurations where the eigenvalues $c_P>c_S>0$ are real. Thus, $c_P$ and $c_S$ correspond to the velocities of compression waves and shear waves, respectively.
	
	\paragraph{Plane waves}
	
	We assume furthermore that the displacement field is invariant along a direction, say $\bm{e}_2$, so that $\bm{u}$ does not depend on $y$. In this case, the vector of unknown reduces to $\mathbf{q} = (u_{1,1},u_{2,1},v_1,v_2,g)^\top\!${\,---\,}the second and fourth rows of \eqref{SystCons}-\eqref{SystConsFlux} are zero{\,---\,}and the flux $\mathbf{g}$ along $y$ is zero. The Jacobian matrix $\mathbf{f}'(\mathbf{q})$ is obtained from \eqref{SystConsJacobiX} by removing the second and fourth rows, as well as the second and fourth columns. Doing so, two zero eigenvalues are removed from the spectrum, which reduces to $\lbrace -c_P, c_P, -c_S, c_S,0\rbrace$.
	
\begin{figure}
		\begin{minipage}{0.52\textwidth}
			\centering
			\includegraphics{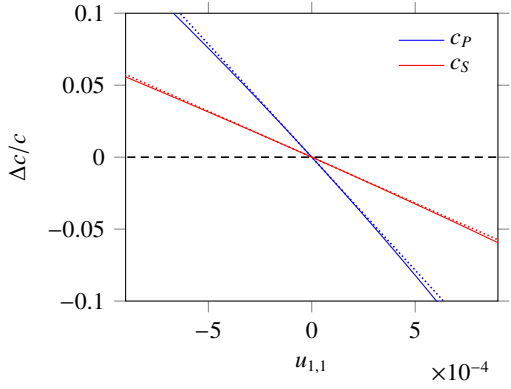}
		\end{minipage}
		\begin{minipage}{0.48\textwidth}
			\caption{Murnaghan hyperelasticity ($g\equiv 0$). Relative variation $\Delta c/c$ of the speeds $c_P$ and $c_S$ of compressional and shear waves with respect to the case of Hooke's law. The variation $\Delta c/c$ is represented with respect to the compression strain $u_{1,1}$, and the shear strain is $u_{2,1} = 0$. The dotted lines mark Taylor series approximations \eqref{SoundSpeedTaylor}. \label{fig:MurnaghanSound}}
		\end{minipage}
	\end{figure}

	We consider the case of Murnaghan material $g\equiv 0$, with the parameters in Table~\ref{tab:ParamMurnag}. The latter, found in \cite{payan09}, have been measured on concrete. Let us introduce the relative variation $\Delta c/c$ of the sound velocities $c_P$ and $c_S$ with respect to the case of Hooke's law, where $c_P = \sqrt{(\lambda+2\mu)/\rho_0} \approx 4458$~m/s and $c_S = \sqrt{\mu/\rho_0} \approx 2700$~m/s. Fig.~\ref{fig:MurnaghanSound} displays the evolution of $\Delta c/c$ with respect to the compression strain $u_{1,1}$, when the shear strain $u_{2,1}$ is set to zero. One observes that the variations of $c_P$ with respect to $u_{1,1}$ are much larger than the variations of $c_S$. This is confirmed by the Taylor series approximations
	\begin{equation}
		{\addtolength{\jot}{0.2em}
		\begin{aligned}
			(\Delta c/c)_P &= \left(\frac{3}{2} + \frac{\mathfrak{l}+2\mathfrak{m}}{\lambda+2\mu}\right) u_{1,1} + O({u_{1,1}}^2) + O({u_{2,1}}^2) \approx {-157}\, u_{1,1}\, , \\
			(\Delta c/c)_S &= \left(\frac{\lambda + 2\mu}{2\mu} + \frac{\mathfrak{m}}{2\mu}\right) u_{1,1} + O({u_{1,1}}^2) + O({u_{2,1}}^2) \approx {-63.8}\, u_{1,1}\, ,
		\end{aligned}}
		\label{SoundSpeedTaylor}
	\end{equation}
	represented as dotted lines in Fig.~\ref{fig:MurnaghanSound} (the magnitude of $u_{1,1}$ in the figure has been chosen for graphical reasons). These approximations show also that the shear strain $u_{2,1}$ has much less influence than the compression strain $u_{1,1}$ on the variations of the sound velocities.
	
	\begin{table}
		\centering
		\caption{Physical parameters of concrete. \label{tab:ParamMurnag}}
		
		{\renewcommand{\arraystretch}{1.2}
			\renewcommand{\tabcolsep}{0.9em}
			\begin{tabular}{ccccccc}
				\toprule
				$\rho_0$ (kg\,m\textsuperscript{$-3$}) & $\lambda$ (GPa) & $\mu$ (GPa) & $\mathfrak{l}$ (GPa) & $\mathfrak{m}$ (GPa) & $\gamma$ (J\,m\,\textsuperscript{$-3$}) & $\tau_1$ (J\,m\textsuperscript{$-3$}\,s) \\
				$2400$ & $12.7$ & $17.5$ & $-3007$ & $-2283$ & $4.0\times 10^{-2}$ & $2.0\times10^{-6}$\\	
				\bottomrule	
		\end{tabular}}
	\end{table}
	
	\section{Numerical resolution}\label{sec:Num}
	
	\subsection{Numerical strategy}\label{subsec:Numstrat}
	
	In the examples presented later on, the physical domain is assumed unbounded. We consider a finite numerical domain $[ 0, L_x ]\times [ 0, L_y ]$. It is discretized using a regular grid in space with mesh size $\Delta x$ in the $x$ direction, and $\Delta y$ in the $y$ direction. The coordinates of the nodes are $(x_i,y_j) = (i\, \Delta x,j\, \Delta y)$, where $0\leqslant i\leqslant N_x$ and $0\leqslant j\leqslant N_y$. The total number of nodes is $(N_x+1)\times (N_y+1)$, where $N_x = L_x/\Delta x$ and $N_y = L_y/\Delta y$. A variable time step $\Delta t = t_{n+1} - t_n$ is introduced. Therefore, $\mathbf{q}(x_i,y_j, t_n)$ denotes the solution to \eqref{SystCons} at the grid node $(i, j)$ and at the $n$th time step. Numerical approximations of the solution are denoted by $\mathbf{q}_{i,j}^n \simeq \mathbf{q}(x_i,y_j, t_n)$.
	
	The non-homogeneous system of balance laws
	\eqref{SystCons} is integrated explicitly in time:
	\begin{equation}
		\mathbf{q}_{i,j}^{n+1} = \mathbf{q}_{i,j}^n + \Delta \mathbf{q}_\text{FV}^n + \Delta t\left( \mathbf{r}(\mathbf{q}_{i,j}^n) + \mathbf{s}_{i,j}^n \right) ,
		\label{SchemaExplCons}
	\end{equation}
	where the approximation $\mathbf{s}_{i,j}^n$ of the source term $\mathbf{s}$ is specified later on.
	The increment $\Delta \mathbf{q}_\text{FV}^n$ is deduced from the integration of $\partial_t\mathbf{q}+\partial_x\mathbf{f}(\mathbf{q})+\partial_y\mathbf{g}(\mathbf{q})=\mathbf{0}$ over one time step. Usually, one has $\Delta \mathbf{q}_\text{FV}^n = (\mathcal{H}_x + \mathcal{H}_y - 2)\, \mathbf{q}_{i,j}^n$,
	where the discrete operators
	\begin{equation}
		\begin{aligned}
			\mathcal{H}_x \mathbf{q}_{i,j}^n &= \mathbf{q}_{i,j}^n -\frac{\Delta t}{\Delta x}\left(\mathbf{f}_{i+1/2,j}^n - \mathbf{f}_{i-1/2,j}^n\right) , \\
			\mathcal{H}_y \mathbf{q}_{i,j}^n &= \mathbf{q}_{i,j}^n - \frac{\Delta t}{\Delta y}\left(\mathbf{g}_{i,j+1/2}^n - \mathbf{g}_{i,j-1/2}^n\right) ,
		\end{aligned}
		\label{DimSplit}
	\end{equation}
	involve the fluxes $\mathbf{f}_{i+1/2,j}^n$, $\mathbf{g}_{i+1/2,j}^n$ of a 2D finite-volume scheme \cite{leveque02}.
	Here, a second-order symmetric dimensional splitting \cite{strang63} is used instead. That is to say, $\mathcal{H}_x$ and $\mathcal{H}_y$ correspond to the integration of $\partial_t\mathbf{q}+\partial_x\mathbf{f}(\mathbf{q})=\mathbf{0}$ and $\partial_t\mathbf{q}+\partial_y\mathbf{g}(\mathbf{q})=\mathbf{0}$ over one time step, so that \eqref{DimSplit} involves the fluxes $\mathbf{f}_{i+1/2,j}^n$, $\mathbf{g}_{i+1/2,j}^n$ of a 1D finite-volume scheme. The increment $\Delta \mathbf{q}_\text{FV}^n$ is computed according to
	\begin{equation}
		\Delta \mathbf{q}_\text{FV}^n = \frac{1}{2}\left(\mathcal{H}_x\mathcal{H}_y + \mathcal{H}_y\mathcal{H}_x - 2\right) \mathbf{q}_{i,j}^n \, ,
		\label{IncrSplit}
	\end{equation}
	where $\mathcal{H}_x\mathcal{H}_y$ denotes the composition of the operators $\mathcal{H}_x$ and $\mathcal{H}_y$.
	
	The numerical fluxes in \eqref{DimSplit}-\eqref{IncrSplit} are computed according to the flux-limiter method \cite{leveque02,berjamin18b} described in the next subsection. This finite-volume scheme is well-suited for nonlinear wave propagation and second-order accurate. The operators $\mathcal{H}_x$ and $\mathcal{H}_y$ are stable under the Courant--Friedrichs--Lewy (CFL) condition
	\begin{equation}
		\text{Co} = \max_{\substack{0\leqslant i\leqslant N_x \\ 0\leqslant j\leqslant N_y}} \max \left\lbrace \varrho_{\mathbf{f}'}(\mathbf{q}_{i,j}^n)\frac{\Delta t}{\Delta x},\, \varrho_{\mathbf{g}'}(\mathbf{q}_{i,j}^n) \frac{\Delta t}{\Delta y} \right\rbrace \leqslant 1 \, ,
		\label{CFL2D}
	\end{equation}
	where $\text{Co}$ is the maximum Courant number in the $x$ and $y$ directions. The spectral radius $\varrho_{\mathbf{f}'}(\mathbf{q})$ of $\mathbf{f}'(\mathbf{q})$ corresponds to $c_P$ (expression detailed in the \ref{app:Eigenvects}), ditto the spectral radius $\varrho_{\mathbf{g}'}(\mathbf{q})$ of $\mathbf{g}'(\mathbf{q})$.
	The stability of the scheme \eqref{SchemaExplCons} is also restricted by the spectral radius of the Jacobian matrix $\mathbf{r}'(\mathbf{q})$. As in 1D \cite{berjamin18b}, the stability limits imply that the scheme \eqref{SchemaExplCons} is stable under the classical CFL condition \eqref{CFL2D}. Hence, given a spatial discretization and a Courant number $\text{Co} \leqslant 1$, the value of the time step $\Delta t$ is imposed by \eqref{CFL2D}.
	
	\subsection{Flux limiter}\label{subsec:PropaPart}

	We describe now the flux-limiter scheme \cite{leveque02,berjamin18b}. Since the computation of the numerical fluxes in the $x$ and $y$ directions is similar, only the numerical flux $\mathbf{f}_{i+1/2,j}^n$ in the $x$ direction is detailed here. To do so, we introduce the Jacobian matrix
	\begin{equation}
		\textstyle
		\mathbf{A}_{i+1/2,j} = \mathbf{f}' \left(\frac{1}{2} (\mathbf{q}_{i,j}^n + \mathbf{q}_{i+1,j}^n)\right) 
		\label{AverageMat}
	\end{equation}
	at the arithmetic mean of the grid node values in the $x$ direction.
	The jump of the numerical solution $\mathbf{q}_{i+1,j}^n - \mathbf{q}_{i,j}^n$ along $x$ is decomposed in the basis of right eigenvectors $\lbrace \mathbf{p}^k_{i+1/2,j}, k = 1,\dots ,7\rbrace$ of $\mathbf{A}_{i+1/2,j}$,
	\begin{equation}
		\mathbf{q}_{i+1,j}^n-\mathbf{q}_{i,j}^n = \sum_{k=1}^7 \alpha^k_{i+1/2,j}\, \mathbf{p}^k_{i+1/2,j} = \sum_{k=1}^7 \bm{\mathcal{W}}^k_{i+1/2,j} \, ,
		\label{DecompJump}
	\end{equation}
	which correspond to the eigenvalues $\lbrace -c_P,c_P,-c_S,c_S, 0,0,0\rbrace$ (cf. detailed expressions in the \ref{app:Eigenvects}).
	
	The numerical flux in \eqref{DimSplit} is the sum of a first-order flux and a second-order limited correction, $\mathbf{f}_{i+1/2,j}^n = \mathbf{f}_{i+1/2,j}^L + \mathbf{f}_{i+1/2,j}^H$, where
	\begin{equation}
		{\addtolength{\jot}{0.2em}
		\begin{aligned}
		\mathbf{f}_{i+1/2,j}^L &= \frac{1}{2}\left( \mathbf{f}(\mathbf{q}_{i,j}^n) +
			\mathbf{f}(\mathbf{q}_{i+1,j}^n)\right) - \frac{1}{2} c_P \left(\bm{\mathcal{W}}^1_{i+1/2,j} + \bm{\mathcal{W}}^2_{i+1/2,j}\right)
			- \frac{1}{2} c_S \left(\bm{\mathcal{W}}^3_{i+1/2,j} + \bm{\mathcal{W}}^4_{i+1/2,j}\right) ,\\
			\mathbf{f}_{i+1/2,j}^H &= \frac{1}{2} c_P\left(1-\frac{\Delta t}{\Delta x}c_P\right) \left(\phi(\theta_{i+1/2,j}^1)\bm{\mathcal{W}}^1_{i+1/2,j} + \phi(\theta_{i+1/2,j}^2)\bm{\mathcal{W}}^2_{i+1/2,j}\right)\\
			&\hspace{0.2em} + \frac{1}{2} c_S\left(1-\frac{\Delta t}{\Delta x}c_S\right) \left(\phi(\theta_{i+1/2,j}^3)\bm{\mathcal{W}}^3_{i+1/2,j} + \phi(\theta_{i+1/2,j}^4)\bm{\mathcal{W}}^4_{i+1/2,j}\right) .
		\end{aligned}}
		\label{NumericalFlux}
	\end{equation}
	The coefficients $\theta_{i+1/2,j}^k$ where $k=1, \dots, 4$ express the upwind variation of the jump \eqref{DecompJump} in the $k$th characteristic field,
	\begin{equation}
		\theta_{i+1/2,j}^{1,3} =  \frac{\bm{\mathcal{W}}^{1,3}_{i+3/2,j}\cdot \bm{\mathcal{W}}^{1,3}_{i+1/2,j}}{\bm{\mathcal{W}}^{1,3}_{i+1/2,j}\cdot \bm{\mathcal{W}}^{1,3}_{i+1/2,j}}
		\, ,\qquad 
		\theta_{i+1/2,j}^{2,4} = \frac{\bm{\mathcal{W}}^{2,4}_{i-1/2,j}\cdot \bm{\mathcal{W}}^{2,4}_{i+1/2,j}}{\bm{\mathcal{W}}^{2,4}_{i+1/2,j}\cdot \bm{\mathcal{W}}^{2,4}_{i+1/2,j}} \, ,
		\label{NumericalFluxThetas}
	\end{equation}
	and $\phi$ denotes the minmod limiter function $\phi(\theta) = \max\lbrace 0,\min\lbrace 1,\theta\rbrace\rbrace$. As such, the weights $\phi(\theta_{i+1/2,j}^k)$ are designed to avoid spurious oscillations in the numerical solution. Since the eigenvalues indexed by $k=5,\dots , 7$ in the decomposition of the jump \eqref{DecompJump} are zero, the corresponding terms $\bm{\mathcal{W}}^k_{i+1/2,j}$ do not appear in the numerical flux \eqref{NumericalFlux}.
	
	To carry out one iteration in time \eqref{SchemaExplCons}-\eqref{IncrSplit} at some grid node $(i, j)$, the numerical values of $\mathbf{q}$ at the grid nodes $(i-2,\dots ,i+2)\times (j-2,\dots ,j+2)$ are required \eqref{NumericalFlux}. Therefore, two columns and two rows of ``ghost cells'' are added on the left, the right, the top, and the bottom of the numerical domain. If not specified differently, a zero-order extrapolation of the numerical values is used to update the ghost cell values at each step of \eqref{IncrSplit}. This procedure is detailed in Section~21.8 of \cite{leveque02}, and is used here to simulate outflow boundary conditions (i.e., an infinite physical domain).
	
	
	\section{Numerical experiments}\label{sec:NumExp}
	
	In the following numerical examples, the Courant number \eqref{CFL2D} is set to $\text{Co} = 0.9$. If not specified otherwise, the physical parameters are given in Table~\ref{tab:ParamMurnag}. The parameters $\gamma$, $\tau_1$ have been chosen so as to obtain significant effects of the softening at the scale of the simulation. The numerical domain is defined by $L_x = L_y = 0.4~\text{m}$, and is discretized using $N_x = N_y = 800$ points in each direction.
	
	\subsection{Murnaghan hyperelasticity}
	
	The first example focuses on nonlinear elastodynamics, i.e., no softening occurs in the material. In \eqref{SystCons}-\eqref{SystConsFlux}, the source term and the relaxation function are removed ($\mathbf{s} = \mathbf{0}$, $\tau_1 \to {+\infty}$). We consider a Riemann problem with initial data $\mathbf{q}(x,y,0)$, where the material is initially undeformed and opposite transverse velocities with amplitude $V$ are applied:
	\begin{equation}
		\mathbf{q}(x,y,0) =
		\left\lbrace
		\begin{aligned}
			& V\, \big(0, 0, 0, 0, \phantom{-\!}\sin\varphi, -\!\cos\varphi, 0\big)^\top & &\text{if}\; x_\varphi < 0 \, , \\
			& V\, \big(0, 0, 0, 0, -\!\sin\varphi, \phantom{-\!}\cos\varphi, 0\big)^\top & &\text{if}\;   x_\varphi > 0 \, .
		\end{aligned}
		\right.
		\label{RiemannProb}
	\end{equation}
	The variable $x_\varphi = (x-x_s) \cos \varphi + (y-y_s) \sin \varphi$ is the $x$-abscissa of a new coordinate system, corresponding to a rotation by an angle $\varphi$ and a translation by $(x_s,y_s)$ of the original one.
	Here, the origin is set at $(x_s,y_s) = (L_x,L_y)/2$, the rotation angle is $\varphi = 15^\circ\!$, and the velocity amplitude is $V = 0.1$~m/s. To reduce discretization artifacts due to the oblique discontinuity, the average value of \eqref{RiemannProb} over the cell $[x_{i-1/2},x_{i+1/2}]\times [y_{j-1/2},y_{j+1/2}]$ is initially set at the grid node $(i, j)$.
	
\begin{figure}
		
		\centering
		\vspace{0.2em}
		
		\includegraphics{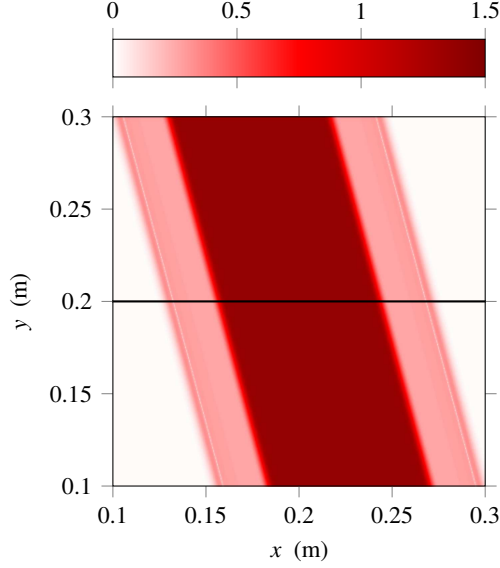}
		
		\caption{Generation of compression waves from pure shear initial data in Murnaghan hyperelasticity ($g\equiv 0$). Map of $W^{1/8}$ at $t = 0.015$~ms, where $W$ is the strain energy density (J/m\textsuperscript{3}) deduced from the numerical solution. The velocity amplitude of the impact problem \eqref{RiemannProb} is $V=0.1$~m/s. \label{fig:RiemannMurnagMap}}
	\end{figure}

\begin{figure}
		\begin{minipage}{0.5\textwidth}
			\centering
			(a)
			\vspace{0.6em}
			
			\includegraphics{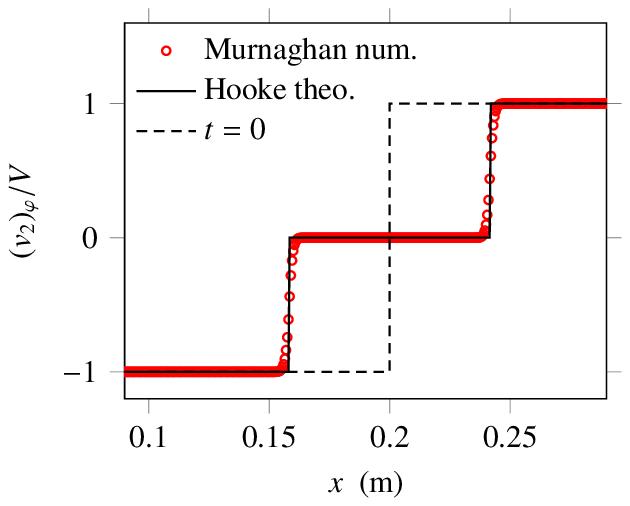}
		\end{minipage}
		\begin{minipage}{0.5\textwidth}
			\centering
			(b)
			
			\includegraphics{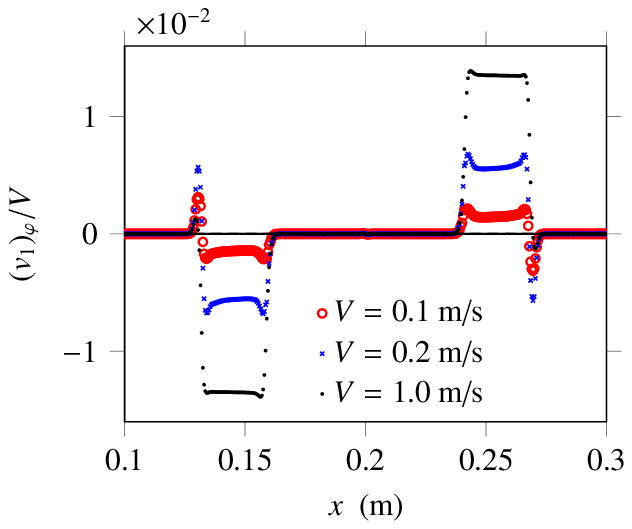}
		\end{minipage}
		
		\caption{Rotated longitudinal and transverse velocities $(v_1)_\varphi$, $(v_2)_\varphi$ along the line $y = 0.2$~m in Fig.~\ref{fig:RiemannMurnagMap}. (a) Normalized velocity $(v_2)_\varphi/V$ in the linear and nonlinear cases for $V = 0.1$~m/s. (b) Normalized velocity $(v_1)_\varphi/V$ in the nonlinear case for various amplitudes $V$. \label{fig:RiemannMurnagCut}}
	\end{figure}

	Figures~\ref{fig:RiemannMurnagMap} and \ref{fig:RiemannMurnagCut} illustrate the coupling between plane shear waves and plane compression waves in hyperelasticity \cite{norris98,favrie14}, contrary to linear elasticity where both types of waves are decoupled. Fig.~\ref{fig:RiemannMurnagMap} displays a map of $W^{1/8}$ at $t = 0.015$~ms, where $W$ is the strain energy \eqref{EnergyStrain} obtained numerically with the above method. Fig.~\ref{fig:RiemannMurnagCut} display the evolution of the rotated longitudinal velocity $(v_1)_\varphi = v_1\cos\varphi + v_2\sin\varphi$ and the rotated transverse velocity $(v_2)_\varphi = -v_1\sin\varphi + v_2\cos\varphi$, along the solid line displayed in Fig.~\ref{fig:RiemannMurnagMap}. In the case of Hooke's law of linear elasticity, the solution to the initial-value problem \eqref{RiemannProb} writes
	\begin{equation}
		\mathbf{q}(x,y,t) =
			\left\lbrace
			{\addtolength{\jot}{0.2em}
			\begin{aligned}
				V\, &\big(0, 0, 0, 0, \phantom{-\!}\sin\varphi, -\!\cos\varphi, 0\big)^\top & &\text{if}\; x_\varphi < -c_S t \, , \\
				\frac{V}{c_S}\, &\big(\! -\!\sin\varphi\cos\varphi, -\!\sin^2\!\varphi, \cos^2\!\varphi, \sin\varphi\cos\varphi, 0, 0, 0\big)^\top & &\text{if}\; {-c_S} t < x_\varphi < c_S t \, , \\
				V\, &\big(0, 0, 0, 0, -\!\sin\varphi, \phantom{-\!}\cos\varphi, 0\big)^\top & &\text{if}\;   c_S t < x_\varphi\, ,
			\end{aligned}}
			\right.
		\label{RiemannLin}
	\end{equation}
	with $c_S = \sqrt{\mu/\rho_0}\,$, and only shear waves propagate (solid line in Fig.~\ref{fig:RiemannMurnagCut}). In the hyperelastic case, Fig.~\ref{fig:RiemannMurnagCut}b shows that shear waves are generated from the initial data \eqref{RiemannProb}, but faster compression waves are also generated. This amplitude-dependent nonlinear effect is  better observed at large amplitudes. The small-amplitude perturbations in the numerical solution are caused by the discretization of the oblique discontinuity \eqref{RiemannProb}.
	
	\subsection{Softening}
	
	In this second example, we consider the full system \eqref{SystCons}-\eqref{SystConsFlux}. The material is initially undeformed and at rest, $\mathbf{q}(x,y,0) = \mathbf{0}$, and a volume force $\bm{f}^v$ is used for the forcing $\mathbf{s}$. The volume force is an acoustic point source along $x$ with expression $\bm{f}^v = A^v \sin(2\pi f_c t) \,\delta(x-x_{s})\,\delta(y-y_{s})\, \bm{e}_1$, where $\delta$ is the Dirac delta function, $A^v$ is the amplitude, and $f_c$ is the characteristic frequency. Usually, the increment $\mathbf{s}_{i,j}^n$ in \eqref{SchemaExplCons} is obtained by averaging the source term $\mathbf{s}|_{t=t_n}$ of \eqref{SystCons}-\eqref{SystConsFlux} over the cell $[x_{i-1/2},x_{i+1/2}]\times [y_{j-1/2},y_{j+1/2}]$. Here, we approximate the Dirac deltas by a truncated Gaussian function to avoid strain concentration at the source. Thus,
	\begin{equation}
		\mathbf{s}_{i,j}^n = 
		\frac{A^v}{\rho_0} \sin(2\pi f_c t_n)\, \frac{\exp\big({-(d_{ij}/\sigma_c)^2}\big)}{\pi {\sigma_c}^2 \left(1-\exp\big({-(R/\sigma_c)^2}\big)\right)} \mathbf{1}_{d_{ij}\leqslant R}\;
		(0,0,0,0,1,0,0)^\top ,
		\label{ForceNum}
	\end{equation}
	where $d_{ij}$ is the distance between $(x_i,y_j)$ and $(x_s, y_s)$.
	Denoted by the indicator function $\mathbf{1}_{d_{ij}\leqslant R}$, the function's support is a disk with radius $R = c_P/(7.5\, f_c)$, where $c_P = \sqrt{(\lambda+2\mu)/\rho_0}$ is the speed of linear compression waves. The width parameter of the Gaussian function is chosen such that $\sigma_c = R/2$.
	The point load has amplitude $A^v = 0.5$~kN/m, frequency $f_c = 100$~kHz, and it is located at the nearest grid node of the domain's center: $(x_s,y_s) \simeq (L_x,L_y)/2$.
	The source \eqref{ForceNum} is switched on at $t=0$, and switched off at $t = 0.04$~ms. Two receivers R1-R2 record the numerical solution during the simulation. R1 is located at $(x_r,y_r) = (0.2,0.22)$~m, and R2 is located at $(x_r,y_r) = (0.2,0.27)$~m.
	
	\begin{figure}		
		\begin{minipage}{\textwidth}
			\centering
			(a)
			
			\includegraphics{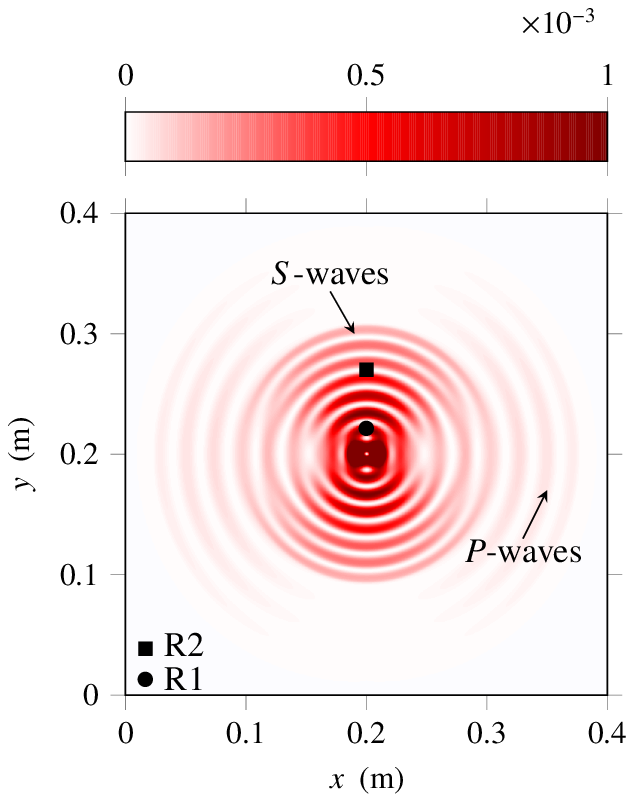}
		\end{minipage}
		
		\begin{minipage}{0.49\textwidth}
			\centering
			(b)
			\vspace{0.2em}
			
			\includegraphics{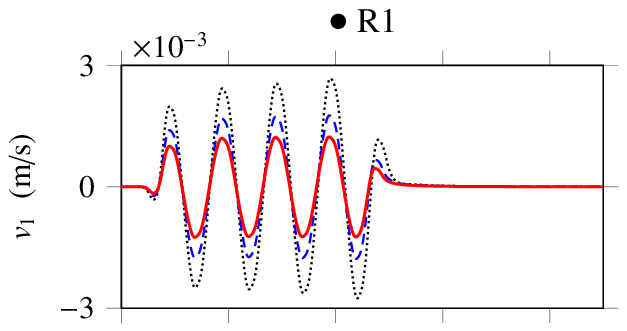}
			\vspace{-2.5em}
			
			\hspace{0.5em}\includegraphics{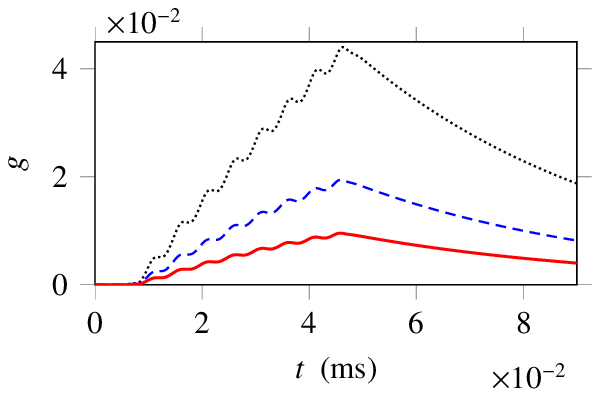}
		\end{minipage}
		\begin{minipage}{0.49\textwidth}
			\centering
			(c)
			\vspace{0.2em}
			
			\includegraphics{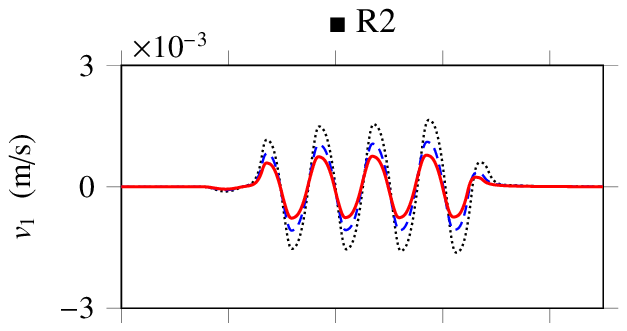}
			\vspace{-2.5em}
			
			\hspace{0.5em}\includegraphics{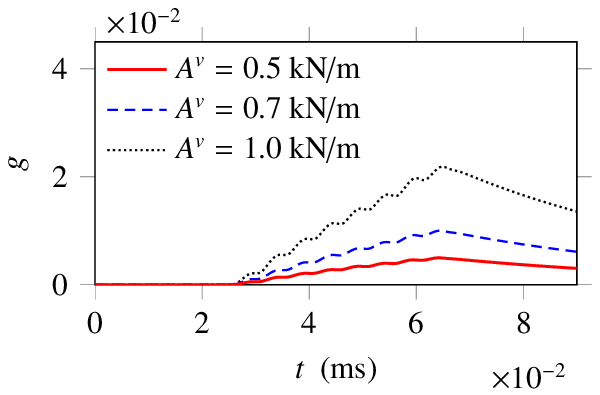}
		\end{minipage}
		
		\caption{
			Softening induced by an acoustic point source \eqref{ForceNum}. (a) Map of the strain energy $\left(1-g\right) W$ in J/m\textsuperscript{3} at $t = 0.04$~ms, where the forcing amplitude is $A^v = 0.5$~kN/m. (b)-(c) Time histories of the velocity component $v_1$ at the position of the receivers R1-R2 (bullet point and square on the map) for several forcing amplitudes (top); Same for the softening variable $g$ (bottom).
			\label{fig:Soft}}
	\end{figure}
	
	Figure~\ref{fig:Soft} illustrates the effect of the softening on the wave propagation. Fig.~\ref{fig:Soft}a displays a map of the strain energy $\left(1-g\right) W$ at the time $t = 0.04$~ms, which shows the propagation of cylindrical waves. Denoted by a bullet point and by a square, the receivers R1-R2 are located in a region of the plane where mainly shear waves propagate. Figs.~\ref{fig:Soft}b-\ref{fig:Soft}c show the effect of the softening at the position of the receivers for several forcing amplitudes $A^v$. One observes that $g$ increases while the wave passes by the receiver, and that it relaxes towards zero afterwards. This softening/recovery process is all the more important as the forcing amplitude is large. The characteristic time of the slow dynamics $\tau_1/\gamma\approx 0.05$~ms corresponds to the characteristic time of the recovery \cite{berjamin17a,berjamin18b}. In Figs.~\ref{fig:Soft}b-\ref{fig:Soft}c, one observes the distortion of the velocity signal during propagation, and its delay due to the increase of $g$. The recorded signals are similar to experimental ones obtained in a longitudinal configuration \cite{remillieux17}.
	
	\subsection{Periodic layered medium}
	
\begin{figure}
			\centering
			\includegraphics{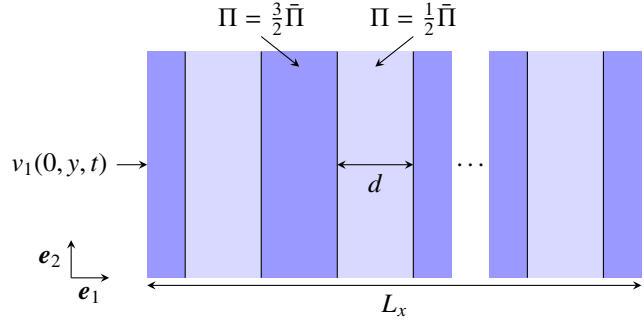}
			
			\caption{Map of the periodic medium's linear material properties $\Pi \in \lbrace\rho_0, \lambda, \mu\rbrace$ over the numerical domain, where $\bar \Pi \in \lbrace\bar \rho_0, \bar \lambda, \bar \mu\rbrace$ denotes the reference values from Table~\ref{tab:ParamMurnag}. \label{fig:Layered}}
		\end{figure}

	Similarly to \cite{fogarty99,leveque02b,leveque03}, a periodic layered medium is considered. The basis vector $\bm{e}_1$ which orientates the $x$-axis is normal to the interfaces. The layers have the same thickness $d = 1$~cm and perfect contact is assumed between neighbor layers (continuity of the displacement and of the stress). As illustrated in Fig.~\ref{fig:Layered}, the linear material properties $\Pi \in \lbrace\rho_0, \lambda, \mu\rbrace$ vary in space according to
		\begin{equation}
			\Pi(x,y) = \left\lbrace
			\begin{aligned}
				& \textstyle\frac{3}{2} \bar \Pi & &\text{if}\quad \lfloor x/d + 0.5\rfloor \equiv 0\; (\text{mod}\; 2) ,\\
				&\textstyle\frac{1}{2} \bar \Pi & &\text{otherwise},
			\end{aligned}\right. 
			\label{SpaceVar}
		\end{equation}
		where $\bar \Pi \in \lbrace\bar \rho_0, \bar \lambda, \bar \mu\rbrace$ denotes the reference parameters given in Table~\ref{tab:ParamMurnag}. The nonlinear material properties $\mathfrak{l}$, $\mathfrak{m}$, $\gamma$, $\tau_1$ do not vary in space.
		A velocity pulse with frequency $f_c = 10$~kHz and amplitude $V = 1$~m/s is imposed at the boundary $x=0$:
		\begin{equation}
			v_1(0,y,t) = V \sin(\pi f_c t)^2\, {\bf 1}_{0\leqslant t \leqslant 1/f_c} \, .
			\label{SourceSoliton}
		\end{equation}
		Since the configuration is invariant along $y$, the problem is one-dimensional. The wavelength of compression waves $c_P^\text{eff}/f_c \approx 39~\text{cm} \gg d$ is deduced from the effective sound speed
		\begin{equation}
			c_P^\text{eff} = \frac{\sqrt{3}}{2} \sqrt{\frac{\bar{\lambda}+2\bar{\mu}}{\bar{\rho}_0}} \approx 3861~\text{m/s}
			\label{C0Eff}
		\end{equation}
		in the linear layered medium.
		
		The numerical domain defined by $L_x = 1$~m is discretized using $N_x = 3000$ points, i.e. each layer is represented by 30~points. The boundary condition \eqref{SourceSoliton} is implemented accordingly to Sec.~7.3.4 of \cite{leveque02} up to $t=1/f_c$. For $t>1/f_c$, periodic boundary conditions are applied (see Sec.~7.1 of \cite{leveque02}). A modification of the numerical method described in Sec.~\ref{sec:Num} is introduced to account for spatially-varying coefficients (\ref{app:LeVeque}).
		
\begin{figure}
			
			\begin{minipage}{0.49\textwidth}
				\centering
				(a)
				
				\includegraphics{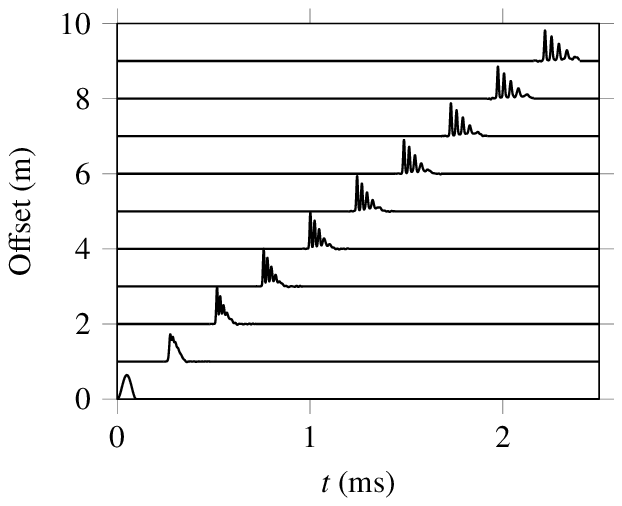}
			\end{minipage}
			\begin{minipage}{0.49\textwidth}
				\centering
				(b)
				
				\includegraphics{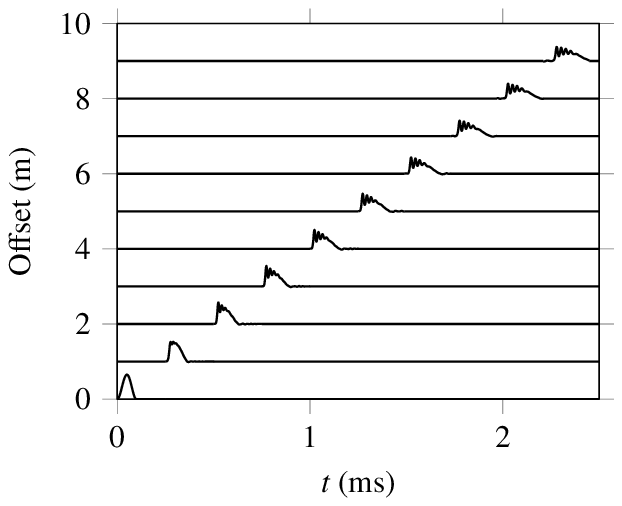}
			\end{minipage}
			\centering
			
			(c)
			
			\includegraphics{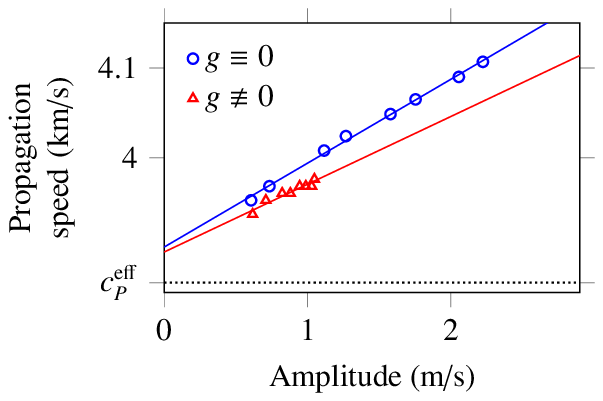}
			
			\caption{(a) Seismograms of the particle velocity $v_1$ for a propagating sinusoidal pulse in a periodic layered Murnaghan material ($g\equiv 0$). (b) Same output in the case of a periodic layered material with softening ($g \not\equiv 0$). (c) In each case, we represent the observed speed of the largest oscillations over the range of propagation distances $[6,8]$~m with respect to their amplitude at the propagation distance $7$~m. \label{fig:Solitons} }
		\end{figure}

		Figure~\ref{fig:Solitons} illustrates the effect of spatially-varying coefficients on the numerical solution up to $t=2.5$~ms. Figs.~\ref{fig:Solitons}a-\ref{fig:Solitons}b display seismograms obtained by unwrapping the velocity signals $v_1(0,y,t)$ recorded at the abscissa $x=0$. The ``offset'' corresponds to the effective propagation distance from the abscissa $x=0$ where the source \eqref{SourceSoliton} is imposed. Fig.~\ref{fig:Solitons}a is obtained with the Murnaghan parameters $\mathfrak{l}$, $\mathfrak{m}$ in Table~\ref{tab:ParamMurnag} while the softening variable $g$ is equal to zero (no softening: $\tau_1\to{+\infty}$). Wavefront steepening is observed, leading to a series of oscillations. Fig.~\ref{fig:Solitons}b shows the same output when the softening variable $g$ evolves according to \eqref{Constitutive}, with the parameters $\gamma = 10^5$~$\text{J}\,\text{m}^{-3}$ and $\tau_1 = 5.0$~$\text{J}\,\text{m}^{-3}\,\text{s}$. The parameters $\gamma$, $\tau_1$ are chosen such that the variable $g$ reaches values of 3\% during the simulation, while the characteristic time $\tau_1/\gamma = 5\times 10^{-5}$~s is the same as in Table~\ref{tab:ParamMurnag}. With respect to the case without softening (Fig.~\ref{fig:Solitons}a), a longer propagation distance is needed until the wave separates into a series of oscillations, and the amplitudes of oscillations are modified.
		
		The speed of the largest oscillations over the range of propagation distances $[6,8]$~m is deduced from Figs.~\ref{fig:Solitons}a-\ref{fig:Solitons}b and reported in Fig.~\ref{fig:Solitons}c, where the abscissa is the amplitude of each oscillation at the propagation distance $7$~m. This procedure is repeated with a smaller input amplitude $V = 0.9$~m/s.
		A nearly linear evolution of the propagation speed with respect to the amplitude is observed, as is the case for solitary waves. At small amplitudes, the propagation speed is close to the effective speed of sound $c_P^\text{eff}$ \eqref{C0Eff}.
		Similar observations are reported in \cite{leveque03} where a different constitutive law is used. Numerically, Fig.~\ref{fig:Solitons}c shows that the softening does not suppress the solitary waves, but it modifies the relationship between their amplitude and their propagation speed.
	
	
	\section{Conclusion}\label{sec:Conclu}
	
	Within the Lagrangian finite-strain theory, the constitutive model used in this study expresses the stress as a function of the Green--Lagrange strain tensor and a softening variable $g$. Also, an evolution equation for $g$ is provided. The system of partial differential equations so-obtained writes as a nonlinear hyperbolic system of balance laws, so that finite-volume methods can be applied. If the softening is neglected, then the material follows Murnaghan's law, where a shear wave excitation induces the propagation of smaller-amplitude compression waves. Otherwise, the propagation of a perturbation is responsible for the softening of the material, which recovers gradually its initial stiffness after excitation has stopped. Several numerical examples are considered along the paper. The latter have been chosen to serve modeling purposes rather than experimental ones, related to the observation of hysteresis and long-time relaxation in geophysics and non-destructive testing. In a periodic layered medium, solitary waves are numerically observed, as known in the case of nonlinear elastodynamics.
	
	Now, let us mention possible future works. In the Lagrangian plane-strain case, the numerical method presented here can be used for various hyperelastic constitutive models, such as neo-Hookean, Mooney--Rivlin, Ogden, etc. \cite{ogdenElast84,holzapfel00}. For such models, it would be interesting to solve analytically the symmetric shear impact problem in Fig.~2 for validation purposes. Variations of the evolution equation of $g$ in \eqref{Constitutive} can be considered as well. The influence of the softening on the smoothness of solutions could be investigated numerically and theoretically. To go further towards realistic configurations, viscoelastic attenuation should be accounted for, e.g. in a similar fashion to \cite{berjamin18a}. Concerning the nonlinear layered medium, it would be interesting to derive the corresponding nonlinear dispersive wave equations by homogenization, as done in \cite{samsonov01,leveque03,andrianov13} and related works. For the development of higher-order methods such as ENO or WENO schemes \cite{shu09}, the eigendecomposition of the fluxes provided in the \ref{app:Eigenvects} is a useful result.
	
	\section*{Acknowledgments}
	
	This work was supported by the interdisciplinary mission of CNRS (INFINITI). The project leading to this publication has received funding from Excellence Initiative of Aix-Marseille University - A*MIDEX, a French ``Investissements d'Avenir'' programme. It has been carried out in the framework of the Labex MEC. Op{\'e}ration r{\'e}alis{\'e}e avec le concours du Programme d'investissements d'avenir du Gouvernement fran{\c c}ais dont la gestion a {\'e}t{\'e} confi{\'e}e {\`a} l'Andra.
	
	
	\appendix
	
	\section{Jacobian matrices of the flux}\label{app:Eigenvects}
	
	\subsection{Expression of the coefficients}
	
	We use the Einstein summation convention with indices in $\lbrace 1 ,2\rbrace$. To encompass both cases with and without geometric nonlinearity in a single equation, we introduce a parameter $\Theta \in \lbrace 0 ,1\rbrace$ such that $\Theta = 1$ corresponds to finite strain and $\Theta = 0$ corresponds to infinitesimal strain. Hence, the coordinates of the strain tensor \eqref{E} are written $E_{ij} = \varepsilon_{ij} + \frac{1}{2}\Theta\, u_{p,i}u_{p,j}$,
	where $\varepsilon_{ij} = \frac{1}{2}(u_{i,j} + u_{j,i})$ are the coordinates of the infinitesimal strain tensor. Moreover, the components of the Piola--Kirchhoff stress tensor \eqref{PK12D} are written $P_{ij} = (1-g) \left(\delta_{im} + \Theta\, u_{i,m}\right) \big(\tilde{\alpha}_0 \delta_{mj} + \tilde{\alpha}_1 E_{mj} \big)$, i.e.
	\begin{equation}
		P_{ij} = (1-g) \left( \tilde{\alpha}_0 \delta_{ij} + \tilde{\alpha}_1 E_{ij} + \Theta \left(\tilde{\alpha}_0 u_{i,j} + \tilde{\alpha}_1 u_{i,m}E_{mj} \right) \right)  ,
		\label{PK12DGen}
	\end{equation}
	where $\tilde{\alpha}_0$, $\tilde{\alpha}_1$ are given in \eqref{PK22Dcoeffs}.
	The coefficients \eqref{SystConsJacobiX} of the Jacobian matrices $\mathbf{f}'(\mathbf{q})$ and $\mathbf{g}'(\mathbf{q})$ satisfy $\rho_0 Q_{ijk\ell} = \partial P_{ij}/\partial u_{k,\ell}$ and $\rho_0 G_{ij} = \partial P_{ij}/\partial g$. In the present case of Murnaghan material with softening \eqref{PK12DGen}, one has
	\begin{equation}
		{\addtolength{\jot}{0.2em}
		\begin{aligned}
			\rho_0 Q_{ijk\ell} &= (1-g) \left(\delta_{ij} \frac{\partial \tilde{\alpha}_0}{\partial u_{k,\ell}} + E_{ij} \frac{\partial \tilde{\alpha}_1}{\partial u_{k,\ell}} + \tilde{\alpha}_1\frac{\partial E_{ij}}{\partial u_{k,\ell}}\right)\\
			&\hspace{-0.8em} + \Theta\, (1-g)\left( u_{i,j} \frac{\partial \tilde{\alpha}_0}{\partial u_{k,\ell}} + \tilde{\alpha}_0\delta_{ik}\delta_{j\ell} + u_{i,m}E_{mj}\frac{\partial \tilde{\alpha}_1}{\partial u_{k,\ell}} + \tilde{\alpha}_1\left( \delta_{ik}E_{j\ell} + u_{i,m}\frac{\partial E_{mj}}{\partial u_{k,\ell}}\right) \right) , \\
			\rho_0 G_{ij} &=  -\left(\tilde{\alpha}_0 \delta_{ij} + \tilde{\alpha}_1 E_{ij} + \Theta \left( \tilde{\alpha}_0 u_{i,j} + \tilde{\alpha}_1  u_{i,m} E_{mj}\right) \right) ,
		\end{aligned}}
		\label{SystConsJacobiCMurnaghan1}
	\end{equation}
	where
	\begingroup
	\addtolength{\jot}{0.2em}
	\begin{align*}
		\frac{\partial E_{ij}}{\partial u_{k,\ell}} &= \frac{1}{2} \big( \delta_{ik} \delta_{j\ell} + \delta_{jk} \delta_{i\ell} \big) + \frac{1}{2} \Theta\, \big(u_{k,i}\delta_{j\ell} + u_{k,j}\delta_{i\ell}\big) \, , &
		\frac{\partial E_{nn}}{\partial u_{k,\ell}} &= \delta_{k\ell} + \Theta\, u_{k,\ell} \, ,\\
		\frac{\partial \tilde{\alpha}_0}{\partial u_{k,\ell}}
		&= \left(\lambda + 2(\mathfrak{l}-\mathfrak{m}) E_{mm}\right) \frac{\partial E_{nn}}{\partial u_{k,\ell}} + 2\mathfrak{m} E_{ij}\frac{\partial E_{ij}}{\partial u_{k,\ell}}  \, , \\
		\frac{\partial \tilde{\alpha}_1}{\partial u_{k,\ell}} &= 2\mathfrak{m} \frac{\partial E_{nn}}{\partial u_{k,\ell}} \, . 
	\end{align*}
	\endgroup
	The case of Hookean solids is recovered if $g \equiv 0$, geometric nonlinearity is neglected ($\Theta = 0$), and the Murnaghan coefficients $\mathfrak{l}$, $\mathfrak{m}$ are zero. In this case, Eq.~\eqref{SystConsJacobiCMurnaghan1} gives $\rho_0 Q_{ijk\ell} = \lambda \delta_{ij}\delta_{k\ell} + \mu (\delta_{ik}\delta_{j\ell} + \delta_{jk}\delta_{i\ell})$.

	\subsection{Eigendecomposition}
	
	We provide an eigendecomposition of the Jacobian matrices $\mathbf{f}'(\mathbf{q})$ and $\mathbf{g}'(\mathbf{q})$ of the fluxes. The hyperelastic case without softening is recovered by removing the last row and the last column of each matrix in the following paragraphs.
	
	\paragraph{Flux along the $x$-axis}
	
	The Jacobian matrix \eqref{SystConsJacobiX} of $\mathbf{f}$ at the linear average \eqref{AverageMat} is diagonalized. Let us write $\mathbf{A}_{i+1/2,j} = \mathbf{P}\mathbf{\Lambda}\mathbf{P}^{-1}$ where $\mathbf{P}$ is an invertible real matrix, and $\mathbf{\Lambda}$ is a diagonal real matrix. The matrix of eigenvalues $\mathbf{\Lambda} = \text{diag}(-c_P,c_P,-c_S,c_S,0,0,0)$ satisfies
	\begin{equation}
		c_{P,S} = \frac{1}{\sqrt{2}} \sqrt{ Q_{1111} + Q_{2121} \pm \sqrt{\left(Q_{1111} - Q_{2121}\right)^2 + 4\, Q_{1121}Q_{2111}}\, } \; ,
		\label{SystDiagoFVals}
	\end{equation}
	where the plus sign gives the expression of $c_P$ (compressional waves), and the minus sign gives the expression of $c_S$ (shear waves).
	The first four right eigenvectors $\mathbf{p}^k_{i+1/2,j}$ of $\mathbf{A}_{i+1/2,j}$ used in \eqref{DecompJump}-\eqref{NumericalFlux} are the first four columns of $\mathbf{P}$, where
	\begin{equation}
		\mathbf{P} =
		{\renewcommand{\arraystretch}{1.2}
		\begin{pmatrix}
			p_{11}       & -p_{11}      & p_{13}       & -p_{13}      & p_{15} & p_{16} & p_{17}\\
			0            & 0            & 0            & 0            & p_{25} & p_{26} & p_{27}\\
			p_{31}       & -p_{31}      & p_{33}       & -p_{33}      & 1 & 0 & 0\\
			0            & 0            & 0            & 0            & 0 & 1 & 0\\
			1            & 1            & p_{13}/p_{33}& p_{13}/p_{33}& 0 & 0 & 0\\
			p_{31}/p_{11}& p_{31}/p_{11}& 1            & 1            & 0 & 0 & 0\\
			0            & 0            & 0            & 0            & 0 & 0 & 1
		\end{pmatrix}}
		\, ,
		\label{SystDiagoFGenR}
	\end{equation}
	with the coefficients
	\begingroup
	\addtolength{\jot}{0.2em}
	\begin{align*}
		p_{11} &= {1}/{c_P} \, ,
		&
		p_{31} &= \frac{Q_{2111} / c_P}{(c_P)^2 - Q_{2121}} \, , \\
		p_{13} &= -\frac{(c_P)^2\, Q_{1121} / c_S}{\left((c_P)^2 - Q_{2121}\right) Q_{1111} + Q_{1121} Q_{2111}} \, ,
		&
		p_{33} &= {1}/{c_S} \, , \\
		p_{15} &= \frac{Q_{1112} Q_{2121} - Q_{1121} Q_{2112}}{Q_{1111} Q_{2112} - Q_{1112} Q_{2111}} \, ,
		&
		p_{25} &= \frac{Q_{1121} Q_{2111} - Q_{1111} Q_{2121}}{Q_{1111} Q_{2112} - Q_{1112} Q_{2111}} \, , \\
		p_{16} &= \frac{Q_{1112}Q_{2122} - Q_{1122}Q_{2112}}{Q_{1111}Q_{2112} - Q_{1112}Q_{2111}} \, ,
		&
		p_{26} &= \frac{Q_{1122} Q_{2111} - Q_{1111} Q_{2122}}{Q_{1111} Q_{2112} - Q_{1112} Q_{2111}} \, , \\
		p_{17} &= \frac{G_{21}Q_{1112}-G_{11}Q_{2112}}{Q_{1111}Q_{2112}-Q_{1112}Q_{2111}} \, ,
		&
		p_{27} &= \frac{G_{11}Q_{2111}-G_{21}Q_{1111}}{Q_{1111}Q_{2112}-Q_{1112}Q_{2111}} \, .
	\end{align*}
	\endgroup
	The matrix $\mathbf{P}$ is invertible provided that its determinant is nonzero, i.e. $Q_{1121}Q_{2111} \neq Q_{1111}Q_{2121}$ and $	Q_{1121}Q_{2111} \neq -\frac{1}{4} (Q_{1111} - Q_{2121})^2$.
	Let us consider each equality case:
	\begin{itemize}
		\item if $Q_{1121}Q_{2111} = Q_{1111}Q_{2121}$, then the eigenvalues of $\mathbf{f}'(\mathbf{q})$ satisfy $c_S =0$. Therefore, the reduced system of conservation laws for plane waves propagating along $x$ is not strictly hyperbolic (eigenvalues $\lbrace {-c_P,c_P,-c_S,c_S,0}\rbrace$);
		\item if $Q_{1121}Q_{2111} = -\frac{1}{4} (Q_{1111} - Q_{2121})^2$, then the eigenvalues of $\mathbf{f}'(\mathbf{q})$ satisfy $c_P = c_S$, which is impossible for the same reason.
	\end{itemize}
	Therefore, the previous eigendecomposition is valid over the domain of strict hyperbolicity. 
	The first four left eigenvectors $\mathbf{l}_{i+1/2,j}^k$ of $\mathbf{A}_{i+1/2,j}$ are the first four rows of $\mathbf{P}^{-1}$, where
	\begin{equation}
		\mathbf{P}^{-1} =
		{\renewcommand{\arraystretch}{1.2}
		\begin{pmatrix}
			q_{11} & q_{12} & q_{13} & q_{14} & q_{15} & q_{16} & q_{17}\\
			-q_{11} & -q_{12} & -q_{13} & -q_{14} & q_{15} & q_{16} & -q_{17}\\
			-q_{41} & -q_{42} & -q_{43} & -q_{44} & q_{45} & q_{46} & -q_{47}\\
			q_{41} & q_{42} & q_{43} & q_{44} & q_{45} & q_{46} & q_{47}\\
			0 & 1/p_{25} & 0 & -p_{26}/p_{25} & 0 & 0 & -p_{27}/p_{25}\\
			0 & 0 & 0 & 1 & 0 & 0 & 0\\
			0 & 0 & 0 & 0 & 0 & 0 & 1
		\end {pmatrix}}
		\, ,
		\label{SystDiagoFGenInv}
	\end{equation}
	with the coefficients
	\begingroup
	\addtolength{\jot}{0.2em}
	\begin{align*}
		q_{11} &= \frac{1}{2}{\frac {p_{33}}{p_{11} p_{33} - p_{13} p_{31}}}
		\, , &
		q_{41} &= \frac{1}{2}{\frac {p_{31}}{p_{11} p_{33} - p_{13} p_{31}}} \, , \\
		q_{12} &= \frac{1}{2}{\frac {p_{13} -p_{15} p_{33}}{p_{25} \left( p_{11} p_{33} - p_{13} p_{31} \right) }}
		\, , &
		q_{42} &= \frac{1}{2}{\frac 	{p_{11} -p_{15} p_{31}}{p_{25}  \left( p_{11} p_{33} - p_{13} p_{31} \right) }} \, , \\
		q_{13} &= -\frac{1}{2}{\frac {p_{13}}{p_{11} p_{33} - p_{13} p_{31}}} \, , &
		q_{43} &= -\frac{1}{2}{\frac {p_{11}}{p_{11} p_{33} - p_{13} p_{31}}} \, ,\\
		q_{14} &= \frac{1}{2}{\frac {\left(p_{15} p_{26} -p_{16} p_{25}\right) p_{33} - p_{13} p_{26}}{p_{25}\left( p_{11} p_{33} - p_{13} p_{31} \right) }} \, , &
		q_{44} &= \frac{1}{2}{\frac {\left(p_{15}p_{26} - p_{16}p_{25}\right) p_{31} - p_{11}p_{26}}{p_{25} \left( p_{11} p_{33} - p_{13} p_{31} \right) }} \, ,\\
		q_{15} &= \frac{1}{2}{\frac {p_{11}p_{33}}{p_{11} p_{33} - p_{13} p_{31}}} \, , &
		q_{45} &= -\frac{1}{2}{\frac {p_{31}p_{33}}{p_{11} p_{33} - p_{13} p_{31}}} \, ,\\
		q_{16} &= -\frac{1}{2}{\frac {p_{13}p_{11}}{p_{11} p_{33} - p_{13} p_{31}}}
		\, , &
		q_{46} &= \frac{1}{2}{\frac {p_{11}p_{33}}{p_{11} p_{33} - p_{13} p_{31}}} \, , \\
		q_{17} &= \frac{1}{2}\frac{\left(p_{15}p_{27}-p_{17}p_{25}\right)p_{33}-p_{13}p_{27}}{p_{25} \left(p_{11}p_{33}-p_{13}p_{31}\right)} \, , &
		q_{47} &= \frac{1}{2}{\frac {\left(p_{15}p_{27} - p_{17}p_{25}\right) p_{31} - p_{11}p_{27}}{p_{25} \left( p_{11} p_{33} - p_{13} p_{31} \right) }} \, .
	\end{align*}
	\endgroup
	The coefficients $\alpha^k_{i+1/2,j}$ in \eqref{DecompJump} are equal to the scalar products $\alpha^k_{i+1/2,j} = \mathbf{l}^k_{i+1/2,j}\cdot \left(\mathbf{q}_{i+1,j}^n - \mathbf{q}_{i,j}^n\right)$.
	
	\paragraph{Flux along the $y$-axis}
	
	Similarly to \eqref{AverageMat}, we introduce the Jacobian matrix $\mathbf{g}'(\mathbf{q})$ of $\mathbf{g}$ at the linear average $\mathbf{B}_{i,j+1/2} = \mathbf{g}'\left(\frac{1}{2}(\mathbf{q}_{i,j}^n + \mathbf{q}_{i,j+1}^n)\right)$, and provide an eigendecomposition $\mathbf{B}_{i,j+1/2} = \mathbf{P}\mathbf{\Lambda}\mathbf{P}^{-1}$. The matrix of eigenvalues $\mathbf{\Lambda} = \text{diag}(-c_P,c_P,-c_S,c_S,0,0,0)$ satisfies
	\begin{equation}
		c_{P,S} = \frac{1}{\sqrt{2}}\sqrt{ Q_{2222} + Q_{1212} \pm \sqrt{\left(Q_{2222} - Q_{1212}\right)^2 + 4\, Q_{2212}Q_{1222}}\, }\; ,
		\label{SystDiagoGVals}
	\end{equation}
	where the plus and minus signs give the expressions of $c_P$ and $c_S$, respectively.
	With similar notations as \eqref{SystDiagoFGenR}, we have
	\begin{equation}
		\mathbf{P} =
		{\renewcommand{\arraystretch}{1.2}
		\begin{pmatrix}
			0 & 0 & 0 & 0 & p_{15} & p_{16} & p_{17} \\
			p_{21} & -p_{21} & p_{23} & -p_{23} & p_{25} & p_{26} & p_{27} \\
			0 & 0 & 0 & 0 & 0 & 1 & 0\\
			p_{41} & -p_{41} & p_{43} & -p_{43} & 1 & 0 & 0 \\
			p_{21}/p_{41} & p_{21}/p_{41} & 1 & 1 & 0 & 0 & 0\\
			1 & 1 & p_{43}/p_{23} & p_{43}/p_{23} & 0 & 0 & 0\\
			0 & 0 & 0 & 0 & 0 & 0 & 1
		\end{pmatrix}}
		\, ,
		\label{SystDiagoGGenR}
	\end{equation}
	with the coefficients
	\begingroup
	\addtolength{\jot}{0.2em}
	\begin{align*}
		p_{21} &= \frac{(c_S)^2\, Q_{1222} / c_P}{\left((c_P)^2 - Q_{1212}\right) Q_{1212} - Q_{1222}Q_{2212}} \, ,
		&
		p_{41} &= 1/c_P \, , \\
		p_{23} &= 1/c_S \, ,
		&
		p_{43} &= -\frac{Q_{2212} / c_S}{(c_P)^2 - Q_{1212}} \, , \\
		p_{15} &= \frac{Q_{1212} Q_{2222} - Q_{1222} Q_{2212}}{Q_{1211} Q_{2212} - Q_{1212} Q_{2211}} \, ,
		&
		p_{25} &= \frac{Q_{1222} Q_{2211} - Q_{1211} Q_{2222}}{Q_{1211} Q_{2212} - Q_{1212} Q_{2211}} \, , \\
		p_{16} &= \frac{Q_{1212} Q_{2221} - Q_{1221} Q_{2212}}{Q_{1211} Q_{2212} - Q_{1212} Q_{2211}} \, ,
		&
		p_{26} &= \frac{Q_{1221} Q_{2211} - Q_{1211} Q_{2221}}{Q_{1211} Q_{2212} - Q_{1212} Q_{2211}} \, , \\
		p_{17} &= \frac{G_{22}Q_{1212}-G_{12}Q_{2212}}{Q_{1211} Q_{2212} - Q_{1212} Q_{2211}} \, ,
		&
		p_{27} &= \frac{G_{12}Q_{2211}-G_{22}Q_{1211}}{Q_{1211} Q_{2212} - Q_{1212} Q_{2211}} \, .
	\end{align*}
	\endgroup
	A similar analysis shows that $\mathbf{P}$ is invertible over the domain of strict hyperbolicity:
	\begin{equation}
		\mathbf{P}^{-1} =
		{\renewcommand{\arraystretch}{1.2}
		\begin{pmatrix}
			q_{11} & q_{12} & q_{13} & q_{14} & q_{15} & q_{16} & q_{17}\\
			-q_{11} & -q_{12} & -q_{13} & -q_{14} & q_{15} & q_{16} & -q_{17}\\
			-q_{41} & -q_{42} & -q_{43} & -q_{44} & q_{45} & q_{46} & -q_{47}\\
			q_{41} & q_{42} & q_{43} & q_{44} & q_{45} & q_{46} & q_{47}\\
			1/p_{25} & 0 & -p_{16}/p_{15} & 0 & 0 & 0 & -p_{17}/p_{15}\\
			0 & 0 & 1 & 0 & 0 & 0 & 0\\
			0 & 0 & 0 & 0 & 0 & 0 & 1
		\end {pmatrix}}
		\, ,
	\label{SystDiagoGGenRInv}
	\end{equation}
	with the coefficients
	\begingroup
	\addtolength{\jot}{0.2em}
	\begin{align*}
		q_{11} &= \frac{1}{2}{\frac {p_{23} - p_{25}p_{43}}{p_{15}\left(p_{21} p_{43} - p_{23} p_{41}\right)}}
		\, , &
		q_{41} &= \frac{1}{2}{\frac {p_{21} - p_{25}p_{41}}{p_{15}\left(p_{21} p_{43} - p_{23} p_{41}\right)}} \, , \\
		q_{12} &= \frac{1}{2}{\frac {p_{43}}{p_{21} p_{43} - p_{23} p_{41}}}
		\, , &
		q_{42} &= \frac{1}{2}{\frac {p_{41}}{p_{21} p_{43} - p_{23} p_{41}}} \, , \\
		q_{13} &= \frac{1}{2}{\frac {\left(p_{16} p_{25} -p_{15} p_{26}\right) p_{43} - p_{16} p_{23}}{p_{15} \left(p_{21} p_{43} - p_{23} p_{41}\right)}} \, , &
		q_{43} &= \frac{1}{2}{\frac {\left(p_{16} p_{25} - p_{15} p_{26}\right) p_{41} - p_{16} p_{21}}{p_{15} \left(p_{21} p_{43} - p_{23} p_{41}\right)}} \, ,\\
		q_{14} &= -\frac{1}{2}{\frac {p_{23}}{p_{21} p_{43} - p_{23} p_{41}}} \, , & 
		q_{44} &= -\frac{1}{2}{\frac {p_{21}}{p_{21} p_{43} - p_{23} p_{41}}} \, ,\\
		q_{15} &= \frac{1}{2}{\frac {p_{43}p_{41}}{p_{21} p_{43} - p_{23} p_{41}}} \, , &
		q_{45} &= -\frac{1}{2}{\frac {p_{41}p_{23}}{p_{21} p_{43} - p_{23} p_{41}}} \, ,\\
		q_{16} &= -\frac{1}{2}{\frac {p_{41}p_{23}}{p_{21} p_{43} - p_{23} p_{41}}}
		\, , & 
		q_{46} &= \frac{1}{2}{\frac {p_{23}p_{21}}{p_{21} p_{43} - p_{23} p_{41}}} \, ,\\
		q_{17} &= \frac{1}{2}{\frac {\left(p_{17} p_{25} -p_{15} p_{27}\right) p_{43} - p_{17} p_{23}}{p_{15} \left(p_{21} p_{43} - p_{23} p_{41}\right)}}
		\, , & 
		q_{47} &= \frac{1}{2}{\frac {\left(p_{17} p_{25} - p_{15} p_{27}\right) p_{41} - p_{17} p_{21}}{p_{15} \left(p_{21} p_{43} - p_{23} p_{41}\right)}} \, .
	\end{align*}
	\endgroup
	
	\section{Spatially-varying coefficients}\label{app:LeVeque}
	
	The problem described in Fig.~\ref{fig:Layered} is one-dimensional. Thus, the variable $y$ and the index $j$ do not appear in the present description of the numerical method, which is a modification of \eqref{NumericalFlux} to account for spatially-varying coefficients \cite{fogarty99,leveque02b,leveque03} (cf. Chap.~9 of \cite{leveque02}). The conservation laws \eqref{SystCons} are rewritten in terms of ${\bf q} = (u_{1,1},\rho_0 v_1,g)^\top\!$. The modified method amounts to choosing the eigenvalues $\lbrace {-c_{P}},c_{P}, 0\rbrace$ and corresponding eigenvectors ${\bf p}_{i+1/2,j}^k$, $k \in \lbrace 1,3\rbrace$ of ${\bf A}_{i+1/2}$ in a downwind fashion. Hence, $c_{P}$ is computed at the abscissa $x_i$ for $k=1$ and at the abscissa $x_{i+1}$ for $k=2$. Since both left-going and right-going waves do not have equal absolute speeds at the cell interface $x_{i+1/2}$, writing the low-order part of the algorithm \eqref{NumericalFlux} as a flux difference is no longer possible: the ``wave-propagation form'' is used instead. To guarantee the continuity of the velocity and of the stress at the material interfaces, a modified expression of the coefficients $\alpha^k_{i+1/2}$ defining the vectors $\bm{\mathcal{W}}^k_{i+1/2}$ is used. In particular, the sum of these vectors does not equal the jump ${\bf q}_{i+1}-{\bf q}_{i}$ anymore, as was the case in a homogeneous medium \eqref{DecompJump}. To avoid numerical instability, a transmission-based limiter is implemented, which modifies the expression \eqref{NumericalFluxThetas} of the coefficients $\theta^k_{i+1/2}$.
	
	
	


\end{document}